\documentclass[english,twocolumn,aps,showpacs,superscriptaddress,amsmath,amssymb,preprintnumbers]{revtex4}
\usepackage[]{fontenc}
\usepackage[latin1]{inputenc}
\usepackage{graphicx}
\usepackage{amssymb}

\newcommand{\pd}[2]{\frac{\partial #1}{\partial #2}}

\makeatletter

\makeatletter

\makeatletter

\makeatletter

\makeatletter

\makeatletter



\makeatother

\makeatother

\makeatother

\makeatother

\makeatother

\usepackage{babel}
\makeatother
\begin{document}

\title{Quasi-Patterns in a Model of Multi-Resonantly Forced Chemical Oscillations}

\author{Jessica Conway and Hermann Riecke}

\affiliation{Engineering Sciences and Applied Mathematics, Northwestern University,
Evanston, IL 60208, USA}
\begin{abstract}
Multi-frequency forcing of systems undergoing a Hopf bifurcation to
spatially homogeneous oscillations is investigated using a complex
Ginzburg-Landau equation that systematically captures weak forcing
functions that simultaneously hit the 1:1-, the 1:2-, and the 1:3-resonance.
Weakly nonlinear analysis shows that generically the forcing function
can be tuned such that resonant triad interactions with weakly damped
modes stabilize subharmonic quasipatterns with 4-fold and 5-fold rotational
symmetry. In simulations starting from random initial conditions domains
of these quasi-patterns compete and yield complex, slowly ordering
patterns. 
\end{abstract}
\maketitle In recent years complex, but ordered spatio-temporal
patterns characterized by multiple length scales have found
considerable interest. In particular on the surface of vertically
vibrated fluid layers (Faraday system) various kinds of fascinating
periodic superlattice patterns and quasi-patterns have been found
experimentally \cite{ChAl92}.  Subsequently, such patterns have also
been observed in optical systems \cite{HeWe99}, in vertically vibrated
fluid convection \cite{RoSc00a}, and on the surface of ferrofluids
driven by time-periodic magnetic fields \cite{KoLe02}. Here we show
that superlattices and quasipatterns should be accessible quite
generally in a different class of systems: resonantly forced systems
undergoing a Hopf bifurcation to spatially homogeneous
oscillations. Paradigmatic for such systems are chemical oscillations
\cite{PeOu97}. In chemical systems patterns with
multiple length scales have so far been obtained only by imposing an
external length through \emph{spatially} periodic illumination
\cite{BeYa03}.

The stability of multi-mode patterns depends on the interaction between
their constitutive Fourier modes. For small angles $\theta$ between
the modes the cross-coupling coefficient $b(\theta)$ is twice as
large as the self-coupling coefficient $b_{0}$. Nearly parallel modes
therefore suppress each other and unless the cross-coupling coefficient
decreases substantially with increasing $\theta$ only stripe-like
patterns are stable. Strong angle dependence can arise if the basic
modes couple to weakly damped, resonating modes \cite{MeTr85,NePo93}. Complex
patterns with different symmetries can then become stable \cite{MaNe89}.

The resonances stabilizing complex patterns have been studied in great
detail in the Faraday system. Their spatio-temporal nature \cite{SiTo00}
allows a very controled tuning through the frequency content of the
driving \cite{PoTo04}. Broadly speaking, there are two mechanisms
by which complex patterns can be stabilized: either by enhancing the
self-damping $b_{0}$ \cite{ZhVi96,NePo93,RuSi07} or by reducing
the cross-coupling coefficient $b(\theta)$ \cite{PoTo04,RuSi07}.

In this paper we exploit spatio-temporal resonances to induce complex
spatial patterns in two-dimensional systems undergoing a Hopf
bifurcation to spatially homogeneous oscillations. To this end we
apply spatially homogeneous, resonant multi-frequency forcing. By including in
the forcing spectrum a frequency component close to twice the Hopf
frequency (1:2-resonance) we excite standing waves with a wavenumber
determined by the detuning between the forcing and the Hopf frequency
\cite{CoFr94}. A second frequency near three times the Hopf frequency
(1:3-resonance) induces a quadratic interaction term, which otherwise
is not allowed in the normal form for Hopf bifurcations. To avoid
transcritical bifurcations off the basic state to hexagonal patterns
we further include a second forcing frequency close to the
1:2-resonance with a slightly different detuning. Within the weakly
nonlinear regime we show that quite independent of the two parameters
characterizing the unforced Hopf bifurcation the forcing function can
be tuned such that instead of the usual stripe, spiral, and
labyrinthine patterns \cite{LiHa04} one obtains superlattices and
quasipatterns.

The systematic, weakly nonlinear description of a weakly forced super-critical
Hopf bifurcation is given by the complex Ginzburg-Landau equation for the
complex oscillation amplitude $A$, which is extended to include near-resonant
components of the forcing function \cite{CoEm92a}, \begin{eqnarray}
\frac{\partial A}{\partial t} & = & (1+i\beta)\nabla^{2}A+\left(\mu+i\sigma-(1+i\alpha)|A|^{2}\right)A\nonumber \\
 &  & +\gamma\left(\cos\chi+\sin\chi e^{i\nu t}\right)A^{*}+\rho e^{i\Phi}A^{*2}.\label{eq:CGL}\end{eqnarray}
 Here $\chi$ measures the relative contributions from the two forcing
components that are close to the 1:2-resonance, which differ in their
frequencies by $\nu$. The detuning between the Hopf frequency and
half the frequency corresponding to the forcing $\gamma\cos\chi$
is given by $\sigma$. The strength and phase of the 1:3-forcing is
given by $\rho$ and $\Phi$, respectively. Nonlinear interactions
of the 1:2-resonant forcing and the 1:3-resonant forcing introduce
an additional forcing near the Hopf frequency itself. To cancel the
resulting additional, inhomogeneous term in (\ref{eq:CGL}) we assume
a further explicit forcing component near the 1:1-resonance. It is
straightforward to derive (\ref{eq:CGL}) from Oregonator-type models
for the photosensitive Belousov-Zhabotinsky reaction \cite{CoRi07a}.

The slight detuning between the two 1:2-forcing components introduces
the explicit, periodic time dependence of the coefficients in (\ref{eq:CGL}).
Using Floquet theory we determine the instability of the basic state
$A=0$ with respect to time-periodic solutions that are phase-locked
to the forcing \cite{CoRi07a}. Due to the dispersion $\beta$ the
detuning $\sigma$ induces phase-locked modes with a non-zero wave
number \cite{CoFr94}. Depending on the forcing parameter $\chi$
and the detuning $\nu$ the mode that destabilizes the basic state
first is either harmonic or subharmonic relative to the period $2\pi/\nu$.

\begin{figure}
\includegraphics[width=5.75cm]{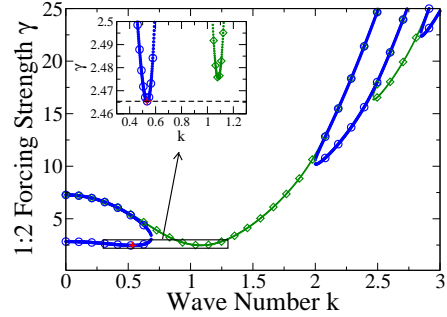}

\caption{Neutral Curves for subharmonic (thick lines) and harmonic modes (thin)
for $\mu=-1$, $\sigma=4$, $\beta=3$, $\chi=0.476718$, and $\nu=4.2$.
\label{fig:Neutral-Curves-for}}
\end{figure}

A typical set of neutral curves $\gamma^{(H,SH)}(k)$ for the harmonic
and subharmonic mode is shown in Fig.\ref{fig:Neutral-Curves-for}.
We focus here on the subharmonic case for which the time-shift symmetry
$t\rightarrow t+2\pi/\nu$ suppresses any transcritical bifurcations
off $A=0$ to hexagons. Weakly damped harmonic modes excited at quadratic
order modify the competition between subharmonic modes of different
orientation \cite{ZhVi96,SiTo00}. Their effect is strongest if the
forcing $\gamma$ is only slightly below the critical forcing strength
$\gamma_{c}^{(H)}$ of the harmonic modes (inset of Fig.\ref{fig:Neutral-Curves-for}).  For this reason we tune the forcing parameter $\chi$ and the detuning $\nu$ so that $\gamma_{c}^{(H)}$ is only slightly above the critical forcing strength $\gamma_{c}^{(SH)}$ of the subharmonic modes.
In this paper we focus on the enhancement of the self-damping $b_{0}$
of the subharmonic modes and choose the forcing function such that
the critical wavenumber $k_{c}^{(H)}$ of the harmonic mode is close
to twice that of the subharmonic mode, $K\equiv k_{c}^{(H)}/k_{c}^{(SH)}\simeq2$.
To reduce the competition between modes subtending a specific angle
$\theta$ a wave-number ratio $K<2$ would be chosen \cite{CoRi07a}.

To compute the interaction between modes of different orientation
within weakly nonlinear analysis we expand the oscillation amplitude
as ($\epsilon\ll1$) \begin{equation}
A=\epsilon\left(A_{1}e^{i\mathbf{k_{1}\cdot}r}+A_{2}e^{i\mathbf{k_{2}\cdot}r}\right)F(t)+h.o.t.\label{eq:expansion}\end{equation}
 For subharmonic patterns $F(t)$ has periodicity $4\pi/\nu$ and
the amplitude equations for $A_{1,2}$ do not contain any quadratic
terms,

\begin{equation}
\frac{dA_{1}}{dt}=\lambda A_{1}-b_{0}{\, A}_{1}\left|A_{1}\right|^{2}-b(\theta){\, A}_{1}\left|A_{2}\right|^{2},\label{eq:amplitude-equation}\end{equation}
 with a similar equation for $A_{2}$.

\begin{figure}
\includegraphics[width=5.75cm]{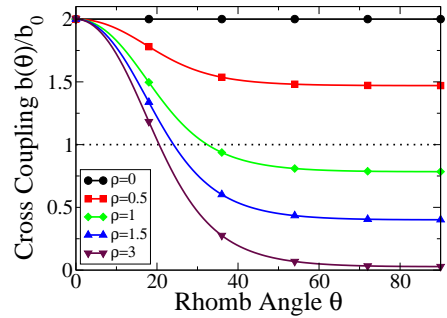}

\caption{Angle-dependence of the effective cross-coupling $b(\theta)/b_{0}$ for $\alpha=-1$ and $\Phi=\frac{3\pi}{4}$. Other parameters as in Fig.\ref{fig:Neutral-Curves-for}.
\label{fig:angle-dependence}}
\end{figure}

Relevant for the pattern selection is the ratio $b(\theta)/b_{0}$.  It
is strongly affected by spatio-temporally resonant triads, which are
induced by the 1:3-forcing $\rho e^{i\Phi}$. The resulting
$\rho-$dependence of $b(\theta)/b_{0}$ is shown in
Fig.\ref{fig:angle-dependence}.  The stability conditions for
rectangular patterns (corresponding to a rhombic arrangement of the
wave vectors) are $b_0>0$ and $|b(\theta)/b_{0}|<1$.  Thus, with
increasing 1:3-forcing $\rho$ a large range of angles arises for which
rectangular patterns are stable, whereas without that forcing only
stripe patterns would be obtained.

Given $b(\theta)$, the linear stability of various types of periodic
super-lattice patterns comprised of three or more modes on a fixed
periodic Fourier lattice can be determined systematically \cite{DiSi97}.
The competition \emph{between} different planforms involves, however,
often a collection of modes that cannot be represented on a single
Fourier lattice. We have determined the relative stability of such
planforms approximately by keeping all modes on the critical circle
that are involved in the pattern competition and find bistability
between a number of different complex patterns \cite{CoRi07a}. This
approach ignores possible side-band instabilities \cite{EcRi01} and
higher-order resonances \cite{HiRi04} and does not account for possible
convergence problems due to small divisors \cite{RuRu03}.

\begin{figure}
\includegraphics[width=5.75cm]{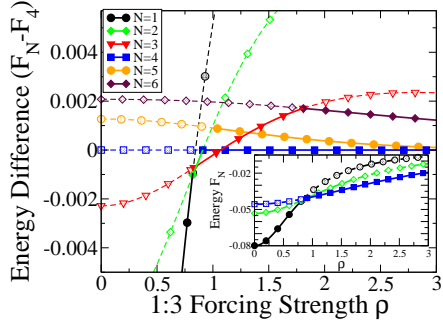}

\caption{Energy of planforms comprised of $N$ equally-spaced Fourier modes.
Parameters as in Fig.\ref{fig:angle-dependence}.  Filled (open) symbols: linearly stable (unstable) planforms.
 \label{fig:Energies}}
\end{figure}

To address the competition between different, simultaneously stable
planforms we exploit the variational character of
(\ref{eq:amplitude-equation}), $\pd{A_j}{T}=-\pd{F_N}{A_j^{*}}$ for
$j=1,\ldots,N$. Fig. \ref{fig:Energies} shows the energies $F_N$ of patterns
comprised of $N$ modes that are equally spaced on the critical circle as a
function of the 1:3-forcing strength $\rho$. When starting from random
initial conditions planforms with lower energy are expected to invade
those with higher energy. Thus, for $\rho\lesssim0.82$ the final state
is expected to consist of stripes, whereas for $\rho\gtrsim1.07$
patterns with four or more modes should dominate.

To test the predictions of our weakly nonlinear analysis we have
performed direct simulations of the complex Ginzburg-Landau equation
(\ref{eq:CGL}).  For small sizes they confirm the linear stability of
periodic patterns comprised of four modes. To investigate the
dependence of the pattern selection on the 1:3-forcing strength $\rho$
we performed simulations in a large square system of linear size
$L\approx 473.39$, which is equivalent to forty wavelengths, for
increasing forcing strengths $\rho$, starting from
\emph{identical} random initial conditions. We have chosen the system
size such that the modes making up hexagons and super-squares have
equal growth rates to bring out clearly how an increase in $\rho$
alone tips the balance from hexagons to 4-fold patterns.  For
$\rho=0.9$ a pattern with hexagonal planform rather than a stripe
structure arises. Because of the reflection symmetry $A\rightarrow-A$
induced by the time-shift symmetry $t\rightarrow t+2\pi/\nu$ domains
with up- and down-hexagons coexist separated by walls containing
narrow layers of triangle patterns \cite{CoRi07a}.

\begin{figure}
\includegraphics[width=5.75cm]{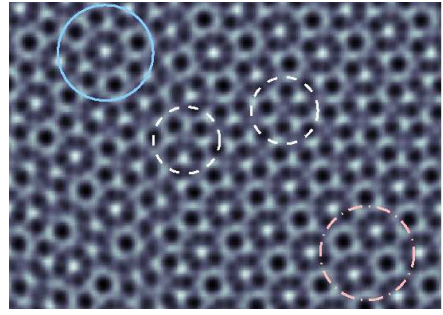}

\caption{Partial view (size $0.5L\times0.35L$) of four-mode
pattern for $\rho=1.2$. Other parameters as in
Fig. \ref{fig:angle-dependence}.
\label{fig:super-squares}}
\end{figure}

\begin{figure}
\includegraphics[width=5.75cm]{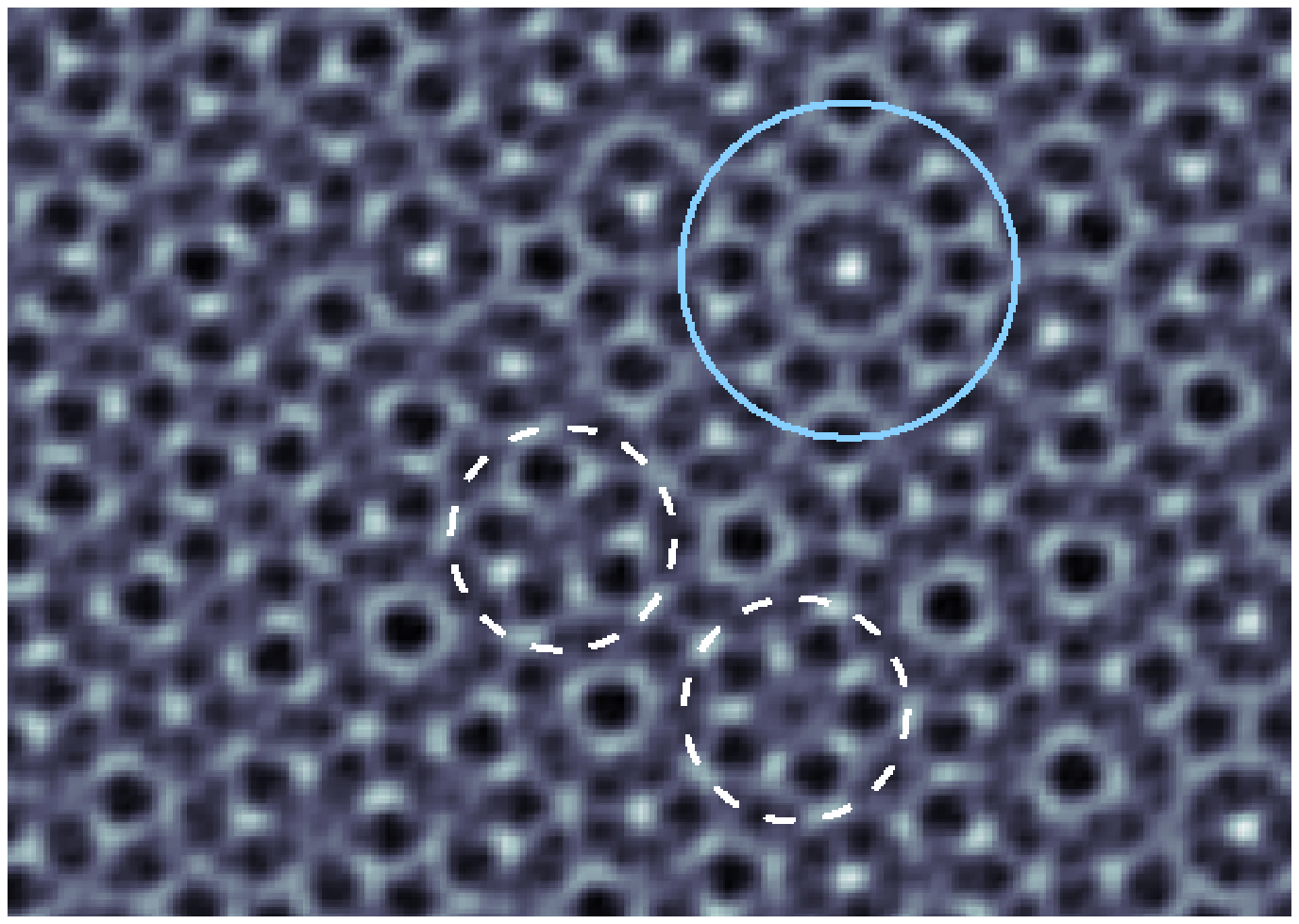}

\caption{Partial view (size $0.5L\times0.35L$) of five-mode
pattern for $\rho=3$. Other parameters as in
Fig.\ref{fig:angle-dependence}.
\label{fig:Five-fold-quasi-patterns}}
\end{figure}

Increasing $\rho$ decreases the $\theta$-range over which modes
suppress each other (cf. Fig.\ref{fig:angle-dependence}) and more
modes persist, as shown in Fig.\ref{fig:super-squares} for $\rho=1.2$.
The pattern exhibits domains of periodic super-lattice patterns with
four-fold rotational symmetry (`super-squares' and `anti-squares'
\cite{DiSi97}, marked by dashed-dotted and dashed circles) as well
as elements with eight-fold rotational symmetry (solid circle).
Increasing the 1:3-forcing to $\rho=3$ increases the number of
persisting modes further and introduces numerous elements with five-
and with ten-fold rotational symmetry (dashed and solid circles) in
Fig. \ref{fig:Five-fold-quasi-patterns}.


\begin{figure}
\includegraphics[width=5.75cm]{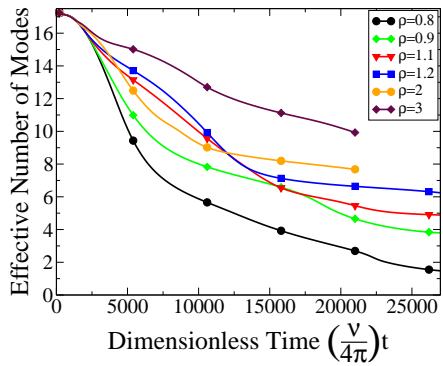}

\caption{Temporal evolution of the relevant number of Fourier modes $e^{S}$
for various values of the 1:3-forcing strength $\rho$. Other parameters
as in Fig.\ref{fig:angle-dependence}. 
\label{fig:entropy}}
\end{figure}

The patterns shown in
Figs.\ref{fig:super-squares},\ref{fig:Five-fold-quasi-patterns} are
still evolving, albeit very slowly. Nevertheless, it is clear that for
$\rho\geq 1.1$ they will not evolve to simple hexagon states.  While
in our simulations for all values of $\rho$ domains of hexagons
appeared for early times, they were replaced for $\rho\geq 1.1$ by
domains of patterns comprised of four or more modes, which have
lower energy.  A condensed view of the temporal evolution of the
patterns for different values of the forcing $\rho$ is given in
Fig.\ref{fig:entropy}. It depicts the evolution of the relevant number
of Fourier modes $e^{S}$, estimated by the spectral pattern entropy
$S\equiv-\sum_{ij}\, p_{ij}\ln p_{ij}$.  Here $p_{ij}$ denotes the
normalized power spectrum. Clearly the number of significant modes
increases with $\rho$, albeit not monotonically at all times.

In conclusion, we have shown that in systems undergoing a Hopf
bifurcation to spatially homogeneous oscillations multi-frequency
forcing can substantially reduce the competition between modes of
different orientation leading to complex multi-mode patterns. By an
appropriate choice of the amplitudes and phases of the forcing
function, which constitute external control parameters, this regime
should be accessible \emph{generically}, essentially independent of
the specifics of the unforced system. Our results should therefore
apply to realistic chemical oscillators \cite{PeOu97,LiHa04,YoHa02}.
From a practical point of view it should be mentioned, however, that
the complex patterns possibly arise only very close to onset. This may
require systems with relatively large aspect ratios and a very precise
tuning of the forcing parameters.

Using direct simulations of the complex Ginzburg-Landau equation we
confirmed that these complex patterns arise from general random initial
conditions. The appropriate, quantitative characterization of the
transients, in which multi-mode structures like super-squares and
anti-squares compete with each other, is still an open problem (cf.
\cite{RiMa06}). Another interesting question is the long-time scaling
of the ordering of such complex structures.

Compared to the Faraday system, the forced Hopf bifurcation considered
here allows an additional level of complexity by going slightly \emph{above}
the Hopf bifurcation. There complex patterns would compete with spatially
homogeneous oscillations. For single-frequency forcing labyrinthine
stripe patterns arise from the oscillations through front instabilities
and stripe nucleation \cite{YoHa02}. It is not known what happens
to this scenario when the stripes are unstable to the more complex
patterns discussed here.

We gratefully acknowledge discussions with A. Rucklidge and M. Silber.
This work was supported by NSF grants DMS-322807 and DMS-0309657.

\bibliographystyle{apsrev}
\bibliography{journal}

\end{document}